\documentclass[final]{aipproc}

\layoutstyle{8x11single}

\usepackage{amsfonts}
\usepackage{amssymb}
\usepackage{graphicx}
\usepackage[latin1]{inputenc}
\newcommand{\R}{\mathcal{R}}
\newcommand{\E}{\mathcal{E}}

\begin{document}

\title{New Phenomenology for Palatini $f(R)$ Theories: Non-singular Universes}


\author{Gonzalo J. Olmo}{
  address={Instituto de Estructura de la Materia, CSIC, Serrano 121, 28006 Madrid, Spain}
}

\begin{abstract}
We study modified theories of gravity of the $f(R)$ type in Palatini formalism. We first consider the stability of atoms when the Palatini gravitational interaction is taken into account in the derivation of the non-relativistic Schrödinger equation. We show that theories with infrared curvature corrections are ruled out by the mere existence of atoms. In particular, we carry out fully perturbative calculations that, for the first time, convincingly rule out the $1/R$ model of Carroll et al. in its Palatini version. We then study the Planck scale corrected quadratic model $f(R)=R+R^2/R_P$ and show that it can avoid the big bang singularity for matter sources which satisfy all the energy conditions. We comment on the mechanisms that cure this singularity and point out that they are closely related to non-perturbative terms also present in the atomic Hamiltonian of infrared corrected models.  
\end{abstract}

\maketitle

\section{Introduction}

The currently accepted cosmological model assumes that the matter and energy contents of the Universe are vastly dominated by two {\it dark} sources whose presence can only be inferred from their gravitational interaction \cite{book1,book2,book3}. And this gravitational interaction is dictated by Einstein's equations. Obviously, if the dynamics were different, the inferred amounts of dark matter and dark energy would be different from those predicted by Einstein's theory. This simple observation has led many authors to consider modifications of the gravitational field equations which could account for the cosmic speed up without the need for sources of dark energy . One of such approaches assumes that the gravity Lagrangian could be some to-be-determined function $f(R)$ which departs from General Relativity (GR), $f(R)=R$, in certain regimes (see \cite{Capozziello:2009nq,Sotiriou:2008rp} and references therein). An interesting aspect of considering modified theories of gravity with non-linear curvature corrections in the Lagrangian is that the derivation of the field equations requires to specify whether the connection is dependent or independent of the metric. The latter case is known as metric-affine formalism, and leads to field equations inequivalent to those found in the more familiar metric formalism, in which the connection is assumed to be compatible with the metric, i.e., it is represented by the Christoffel symbols. When in a metric-affine theory the independent connection is not coupled to the matter fields, then the theory is said to be formulated in the Palatini formalism. \\

The purpose of this work is twofold. First we want to show that Palatini $f(R)$ models with correcting terms that dominate at low curvatures are ruled out by the very existence of atoms. We will see that the characteristic cosmic density scale of such models, which sets the scale at which cosmic speedup arises, can be effectively reached in microscopic systems such as atoms. At such density scales, the modified gravitational dynamics induces strong backreaction effects in the metric which affect the atomic Hamiltonian and end up disintegrating the atom. 
This will lead us to the second point. The strong backreaction effects that spoil infrared corrected models turn into a very effective mechanism to remove singularities in ultraviolet modified models. In fact, we will show that in simple models such as $f(R)=R+R^2/R_P$, where $R_P\sim l_P^{-2}$ is of the order of the Planck curvature, the Big Bang singularity of General Relativity can be replaced by a bounce even if the universe is dominated by matter sources which satisfy all the energy conditions. If this theory is formulated in the usual metric formalism, no bouncing solutions exist \cite{Novello:2008ra}.   \\

The content is organized as follows. We first derive the field equations of a generic $f(R)$ Lagrangian and briefly comment on its weak field limit. Second, we discuss the effective Hamiltonian that governs the motion of an electron in a Hydrogen atom and which is subject to the Palatini gravitational interaction. We show that infrared curvature corrections in the Lagrangian generate instabilities which end up disintegrating the atom, which clearly rules out such models. We then show that in ultraviolet corrected models similar mechanisms also appear in critical scenarios such as the very early Universe. The effect in this case is the avoidance of the Big Bang singularity by the emergence of a cosmic bounce due to strong backreaction effects. In these models atoms are as stable as in GR. We end with a brief summary and conclusions.

\section{Field Equations}

Let us begin by defining the action of Palatini theories
\begin{equation}\label{eq:Pal-Action}
S[{g},\Gamma ,\psi_m]=\frac{1}{2\kappa^2}\int d^4
x\sqrt{-{g}}f({R})+S_m[{g}_{\mu \nu},\psi_m]
\end{equation}
Here $f({R})$ is a function of ${R}\equiv{g}^{\mu \nu }R_{\mu \nu }(\Gamma )$, with $R_{\mu \nu }(\Gamma )$ given by
$R_{\mu\nu}(\Gamma )=-\partial_{\mu}
\Gamma^{\lambda}_{\lambda\nu}+\partial_{\lambda}
\Gamma^{\lambda}_{\mu\nu}+\Gamma^{\lambda}_{\mu\rho}\Gamma^{\rho}_{\nu\lambda}-\Gamma^{\lambda}_{\nu\rho}\Gamma^{\rho}_{\mu\lambda}$
where $\Gamma^\lambda _{\mu \nu }$ is the connection. The matter action $S_m$ depends on the matter fields $\psi_m$, the metric $g_{\mu\nu}$, which defines the line element $ds^2=g_{\mu\nu}dx^\mu dx^\nu$, and its first derivatives (Christoffel symbols). The matter action does not depend on the connection $\Gamma^\lambda _{\mu \nu }$, which is seen as an independent field appearing only in the gravitational action (this condition is not essential and can be relaxed at the cost of introducing a non-vanishing torsion). Varying (\ref{eq:Pal-Action}) with respect to the metric $g_{\mu\nu}$ we obtain
\begin{equation}\label{eq:met-var-P}
f_R(R)R_{\mu\nu}(\Gamma)-\frac{1}{2}f(R)g_{\mu\nu}=\kappa ^2T_{\mu
\nu }
\end{equation}
where $f_R(R)\equiv df/dR$. From this equation we see that the scalar $R$ can be solved as an algebraic function of the trace $T$. This follows from the trace of
(\ref{eq:met-var-P})
\begin{equation}\label{eq:trace-P}
f_R(R)R-2f(R)=\kappa ^2T,
\end{equation}
The solution to this algebraic equation will be denoted by $R=\R(T)$. The variation of (\ref{eq:Pal-Action}) with respect to $\Gamma^\lambda _{\mu \nu }$ leads to $\nabla_\rho  \left[\sqrt{-g} f_Rg^{\mu \nu }\right]=0$,
where $f_R\equiv f_R(\R[T])$ is also a function of the matter terms. This equation leads to 
\begin{equation}\label{eq:Gamma-1}
\Gamma^\lambda_{\mu \nu }=\frac{t^{\lambda \rho
}}{2}\left(\partial_\mu t_{\rho \nu }+\partial_\nu
t_{\rho \mu }-\partial_\rho t_{\mu \nu }\right)
\end{equation}
where  $t_{\mu \nu }\equiv \phi g_{\mu \nu }$, and $\phi\equiv \frac{f_R(\R[T])}{f_R(\R[0])}$ is dimensionless and normalized to unity outside of the sources ($T=0$). It is now useful to rewrite (\ref{eq:met-var-P}) adding and subtracting $\frac{f_R}{2}\R(T)g_{\mu\nu}\equiv
\frac{f_R}{2}t^{\alpha\beta}R_{\alpha\beta}(\Gamma)t_{\mu\nu}$ to get 
\begin{equation}\label{eq:G-tmn}
f_RG_{\mu\nu}(t)=\kappa^2T_{\mu\nu}-\frac{[\R f_R-f]}{2\phi}t_{\mu\nu}
\end{equation}
where $G_{\mu\nu}(t)$ is the Einstein tensor associated to $t_{\mu\nu}$. The equations of motion (\ref{eq:G-tmn}) for the auxiliary metric $t_{\mu\nu}$ are (formally) considerably simpler than those for $g_{\mu\nu}$: 
\begin{eqnarray}\label{eq:Gmn}
R_{\mu \nu }(g)-\frac{1}{2}g_{\mu \nu }R(g)&=&\frac{\kappa
^2}{f_R}T_{\mu \nu }-\frac{\R f_R-f}{2f_R}g_{\mu \nu
}-\frac{3}{2(f_R)^2}\left[\partial_\mu f_R\partial_\nu
f_R-\frac{1}{2}g_{\mu \nu }(\partial f_R)^2\right]+ \frac{1}{f_R}\left[\nabla_\mu \nabla_\nu f_R-g_{\mu \nu }\Box
f_R\right] \ .
\end{eqnarray}
Solving for $t_{\mu\nu}$ using the system (\ref{eq:G-tmn}) and then going back to $g_{\mu\nu}$ via the conformal transformation $g_{\mu\nu}=\phi(T)^{-1}t_{\mu\nu}$ is a useful simplification that makes the task of finding solutions much easier. Note that this fortunate circumstance is due to the fact that the conformal transformation completely cancels out the disturbing derivatives on the right hand side of (\ref{eq:Gmn}). Moreover, the conformal relation between the two metrics puts forward the fact that the metric $g_{\mu\nu}$ receives two kinds of contributions: the usual non-local contributions that result from integration over the sources (like in GR), which produce the term $t_{\mu\nu}$, and local contributions due to $\phi(T)$, which depend on the local details of $T$ at each space-time point. This local contribution arises due to the independent character of the connection and, to our knowledge, is not present in any other metric theory of gravity, where the connection is generally assumed to be metric compatible (Levi-Cività connection). \\

The similarity between the field equations of GR and (\ref{eq:G-tmn}) suggests that for weak sources and reasonable choices of Lagrangian $f(R)$ (those that lead to negligible {\it cosmological term} $\frac{\R f_R-f}{2f_R}$), the right hand side of (\ref{eq:G-tmn}) is small and, like in GR, $t_{\mu\nu}$ can be expressed as $t_{\mu\nu}(x)\approx \eta_{\mu\nu}+h_{\mu\nu}(x)$, where $|h_{\mu\nu}(x)|\ll 1$ is given as an integral over the sources (for details of exact calculations see \cite{Olmo07}). For very weak sources such as atoms or elementary particles, we find that the self-contribution to $h_{\mu\nu}\to 0$. On the other hand, if our microscopic system is placed in an external gravitational field whose contribution to $t_{\mu\nu}$ is not negligible, we can always take a coordinate system in which the metric at the boundaries of a {\it box} containing the system (large as compared to the microscopic system but small as compared to the range of variation of the external metric $t_{\mu\nu}$) becomes $\sim \eta_{\mu\nu}$. In both situations, the metric $g_{\mu\nu}$  becomes simply
\begin{equation}\label{eq:phi-eta}
g_{\mu\nu}(x)\approx\phi(T)^{-1}\eta_{\mu\nu}
\end{equation}
where $\phi(T)\to 1$ in the (empty) boundaries of our {\it auxiliary box} but depart from unity in the region supporting the sources (which are assumed to be fields spread over the space). For a detailed discussion of the Newtonian and post-Newtonian limits of $f(R)$ theories see \cite{Olmo05,Olmo07b}.\\

\section{Atomic Instabilities}

It is well known \cite{Parker80PRL,Parker80,P-P82} that the energy levels of a Hydrogen atom falling freely in an external gravitational field (in GR) will be shifted in a very characteristic way due to the interaction of the electron with the curvature of the space-time. Though external fields in Palatini theories of gravity must also lead to this phenomenon, we will focus here on a different aspect. We will study the effect that the local energy-momentum densities have on the non-relativistic limit of the Dirac equation due to the factor $\phi(T)^{-1}$ appearing in (\ref{eq:phi-eta}). The non-relativistic Schrödinger equation that governs the motion of an electron in an external electromagnetic field taking into account the Palatini gravitational interaction is the following (see \cite{Olmo08a,Olmo07} for a detailed derivation and discussion)
\begin{eqnarray}\label{eq:Pauli}
\E\eta&=& \left\{\frac{1}{\tilde{m}+m_0}[(\vec{p}-e\vec{A})^2-e\vec{\sigma}\cdot\vec{B}]+eA_0\right\}\eta \nonumber\\&+&\left\{\frac{1}{\tilde{m}+m_0}\left[i\vec{\sigma}(\vec{\nabla}\Omega\times\vec{\nabla})-2ie(\vec{A}\cdot\vec{\nabla}\Omega)\right.\right. \\&+&\left.\left.\vec{\nabla}^2\Omega-|\vec{\nabla}\Omega|^2+2(\vec{\nabla}\Omega \cdot\vec{\nabla})\right]+(\tilde{m}-m_0)\right\}\eta  \nonumber \ , 
\end{eqnarray}
where $\E$ represents the total energy minus the rest mass energy $m_0$, $\Omega \equiv (3/4)\ln \phi(T)$, $\tilde{m}\equiv m\phi^{-\frac{1}{2}}$, $m\sim m_0$ is the mass appearing in the action of the Dirac spinor, and $\eta$ is the large component of the Dirac spinor  (the positive energy Foldy-Wouthuysen bispinor).

The first line of this equation is very similar to the well-known non-relativistic Schrodinger-Pauli equation (see (\ref{eq:Sch-Pauli}) below). The only difference being the term $1/(\tilde{m}+m_0)$. The second and third lines, however, represent completely new terms generated by the Palatini gravitational interaction. When the gravity Lagrangian is that of GR, $\phi(T)=1$, we recover the Schrodinger-Pauli equation if $m_0$ is identified with ${m}$.\\
To extract physical results from this equation, we proceed as follows. We first solve (\ref{eq:Pauli}) in the case of GR, $f(R)=R, \phi(T)=1$, which is well known. Then we switch to a different gravity Lagrangian (assuming that we have the ability to do that) and study how the system reacts to that change. The reason for this is that in a general $f(R)$ the metric is sensitive to the local $T_{\mu\nu}$ via $\phi(T)^{-1}$, and changes in the metric due to the matter distribution could react back on the matter equations. If the new interaction terms in (\ref{eq:Pauli}) lead to small perturbations, then the initial wavefunctions will be, roughly speaking, stable with perhaps small corrections which could be computed using standard approximation methods. If, on the contrary, the energy associated to the gravitationally-induced terms is large, that would mean that the original configuration is not minimizing the modified Hamiltonian and, therefore, large modifications would be necessary to reach a new equilibrium configuration. Depending on the magnitude of the reaction on the system, we could estimate whether the theory is ruled out or not. \\

\begin{figure}
\includegraphics[angle=0,width=3.0in,clip]{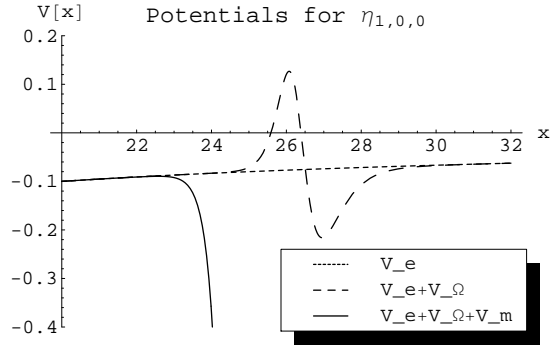}
\caption{Contribution of the different potentials in the ground state. The solid line, which represents the sum of all the potentials, tends to the constant value $-0.134m_0$ (or $-5048$ in the units of the plot).}\label{Fig:ground}
\end{figure}

Let us now focus on the infrared corrected model $f(R)=R-\frac{\mu^4}{R}$, initially proposed in \cite{CDTT04} within the metric formalism and in  \cite{Vol03} in the Palatini version, which is characterized by the low curvature scale $\mu^2$. In this particular model, we find that $\R(T)=-(\kappa^2 T+\sqrt{(\kappa^2 T)^2+12\mu^4})/2$ recovers the GR limit for $|\kappa^2 T|\gg \mu^2$ and tends to a constant $\R\sim \mu^2$ for $|\kappa^2 T|\ll \mu^2$. It also follows that the function $\phi(T)$ is given by
\begin{equation}\label{eq:f_R}
\phi(T)=1-\frac{1}{2[1+\sqrt{1+12/\tau^2}]} \ ,
\end{equation}
where $\tau\equiv -T/T_c$, $T_c\equiv\mu^2/\kappa^2\equiv\rho_\mu $, and $\rho_\mu\sim 10^{-26} \ g/cm^3$ represents the characteristic cosmic density scale of the theory, which triggers the cosmic speedup. It is easy to see that at high densities, as compared to $\rho_\mu$, $\phi(T)\to 3/4$, whereas for $\rho\ll \rho_\mu$ we find $\phi(T)\to 1$. \\
Expressing length units in terms of the Bohr radius ($a_0\sim 0.53\cdot 10^{-10}$m), we find that $\tau=\frac{\rho_e(x)}{\rho_\mu}=10^{24} P_e(x)$, where we have intentionally omitted the nuclear contribution (only relevant at the origin) for simplicity and have used $\rho_e(x)=m P_e(x)$ with $P_e(x)=\eta^\dagger(x)\eta(x)$. This expression for $\tau$ indicates that the electron reaches the characteristic cosmic density, $\tau\sim 1$, in regions where the probability density is near $P_e(x)\sim 10^{-24}$. In ordinary applications, one would say that the chance to find an electron in such regions is negligible, that that region is empty. In our case, however, that scale defines the transition between the high density ($\tau\gg 1$) and the low density ($\tau\ll 1$) regions.\\ In regions of high density, we find that $\phi$ rapidly tends to a constant, $\phi_\infty=3/4$, which leads to $\tilde{m}=2m/\sqrt{3}$ and $\vec{\nabla}\Omega=0$. If we then identify $m\to \sqrt{3}m_0/2$, equation (\ref{eq:Pauli}) reduces to the usual Schrodinger-Pauli equation
\begin{equation}\label{eq:Sch-Pauli}
\E\eta= \left\{\frac{1}{2m_0}[(\vec{p}-e\vec{A})^2-e\vec{\sigma}\cdot\vec{B}]+eA_0\right\}\eta 
\end{equation}
This fact clarifies the introduction of $m_0$ above and guarantees that the solutions of the equation in this region coincide with the standard ones (those of GR). Therefore, the perturbative approach that we have followed is well justified. Let us now see what happens in regions of low density. In those regions,  $\phi(T)$ tends to unity, $\vec{\nabla}\Omega=0$, and $\tilde{m}\to m$ as $\tau\to 0$. The mass factor dividing the kinetic term is now a bit smaller ($m_0>m$) than in the high density region. But the mass difference $\tilde{m}-m_0$ is not zero. This is a remarkable point, because $\tilde{m}-m_0\approx -0.13m_0$ is negative and of order $\sim m_0$, which represents a large contribution to the Hamiltonian. Modifications of the standard solutions are thus expected in this region. To better understand the effect of this term, it is useful to consider the ground state, $\eta_{(1,0,0)}=\frac{e^{-r/a_0}}{\sqrt{\pi a_0^3}}\otimes |\frac{1}{2},s\rangle$, where $|\frac{1}{2},s\rangle$ represents a normalized constant bispinor. In this case, the transition from the high density region to the low density region occurs at $r\approx 26a_0$. In Fig.\ref{Fig:ground} we have plotted the most representative potentials in dimensionless form
\begin{eqnarray}
V_e&=& -\frac{2}{x} \label{eq:Ve}\\
V_m&=& \frac{2m_0c^2a_0^2}{\hbar^2}\left[m\phi^{-\frac{1}{2}}-m_0\right] \label{eq:Vm}\\
V_\Omega&=& \left[\frac{2}{x}\partial_x\Omega+\partial^2_x\Omega-|\partial_x\Omega|^2\right] \label{eq:VO}
\end{eqnarray}
where $V_e$ is the electrostatic potential generated by the proton, lengths are measured in units of the Bohr radius, $x=r/a_0$, and energies in units of $\frac{\hbar^2}{2m_0 a_0^2}\approx 13.6$ eV. Note that $V_\Omega=\vec{\nabla}^2\Omega-|\vec{\nabla}\Omega|^2$ only contains the most important contributions associated to $\Omega$.

In this case, $V_\Omega$ represents a small transient perturbation. The mass difference $V_m$, however, introduces a deep potential well in the outermost parts of the atom that must have important consequences for its stability. Note that this effect is not an artifact of the non-relativistic approximation, since it also occurs in the full relativistic theory (see \cite{Olmo08a} for details). In the initial configuration of the atom, corresponding to GR, the wavefunction of the ground state is concentrated near the origin, where the attractive electric potential is more powerful ($V_e\to-\infty$). As we switch on the $1/R$ theory, a deep potential well of magnitude $\sim-0.13 m_0$ appears in the outer regions of the atom, where $\rho_e(x)\lesssim \rho_\mu$, which makes the ground state unstable and triggers a flux of probability density (via quantum tunneling) to those regions. The half life of Hydrogen subject to this potential can be estimated using time dependent perturbation theory (see \cite{Olmo08a}) yielding
\begin{equation}\label{eq:halflife}
\tau\equiv\frac{\hbar}{\Gamma}\approx 6\cdot 10^3 s
\end{equation}
We thus see that the initial, stable configuration is destroyed in a lapse of time much shorter than the age of the Universe, which is in clear conflict with experiments. \\

Further evidence supporting the instability of the atom is found in the existence of zeros in the atomic wavefunctions in between regions of high density because, obviously, before (and after) reaching $\rho_e(x)=0$ the characteristic scale $\rho_e(x)\sim \rho_\mu$ is crossed. The  first excited state, $\eta_{(2,0,0)}=\frac{1}{\sqrt{8\pi a_0^3}}(1-\frac{r}{2a_0})e^{-r/2a_0}\otimes |\frac{1}{2},s\rangle$, has a zero at $r=2a_0$. The radial derivatives of $\phi(T)$ at that point are very large and lead to very important perturbations which overwhelmingly dominate over any other contribution (see Fig.2). 
\begin{figure}
\includegraphics[angle=0,width=3.0in,clip]{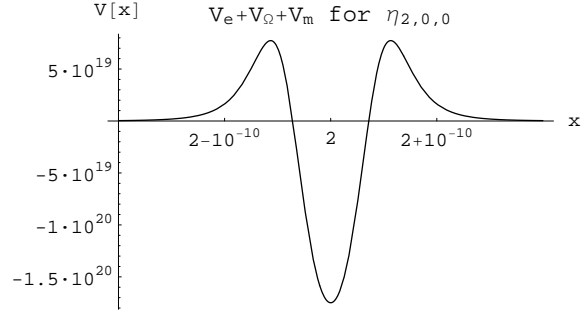}
\caption{The different contributions in this plot are $V_e\sim -1$, $V_m\sim -5\cdot 10^3$, and $V_\Omega\sim \pm 10^{19}$. The y-axis is measured in units of $13.6$eV; the x-axis in units of $a_0$.}\label{fig:2}
\end{figure}

The magnitude of $V_\Omega=\vec{\nabla}^2\Omega-|\vec{\nabla}\Omega|^2$ at $r=2a_0$ oscillates between $10^{20}$ and $-10^{21}$ eV in an interval of only $2\cdot 10^{-10}a_0$. Needless to say that this configuration cannot be stable and that strong changes must take place in the wave function to reduce the energy of the system. Such changes should tend to reduce the magnitude of the density gradients ($\vec{\nabla}\Omega$) to minimize the value of $V_\Omega$, which will likely lead to a rapid transition to the ground state, where $V_\Omega$ is small. One can easily verify that strong gradients $\vec{\nabla}\Omega$ also appear at the zeros of all the $\eta_{n,0,0}$ wavefunctions, which generate  large contributions $V_\Omega$ in those regions. Furthermore, if one considers stationary states with $l\neq 0$, $V_\Omega$ has important contributions not only at the zeros of the radial functions, but also at the zeros of the angular terms. Thus, the pathological behavior described for the spherically symmetric modes gets worse for the $l\neq 0$ states. One thus expects the decay of these states into states with less structure (weaker gradients) such as the ground state,  which will later decay into the continuum.  All this indicates that the existence of bound states, with localized regions of high probability density (where ``high'' means above the scale $\rho_\mu$), are impossible in this theory because of the large gradient contributions $V_\Omega$ and the deep potential well $V_m$. \\

Though we have only analyzed in detail the infrared-corrected model $f(R)=R-\mu^4/R$, the instabilities associated to the potential well $\tilde{m}-m_0$ and the zeros of the wavefunction must be present in all gravity models sensitive to low curvature/energy-density scales. Therefore, such models are ruled out by the mere existence of atoms. Note that models with high curvature corrections, such as $f(R)=R+R^2/R_P$, do not exhibit any instabilities of those affecting infrared corrected models. In this example, we find that $\phi=1-2\kappa^2T/R_P$, which implies that all gradients appearing in $V_\Omega$ are suppressed by the Planck density $\rho_P=R_P/\kappa^2$ and also that $\tilde m= m_0(1+O[\rho/\rho_P])\approx m_0$. The effects of the potentials induced by the $\phi(T)^{-1}$ factor in front of the metric are, therefore, much smaller than the  corrections due to the Newtonian potential. Thus, in such models atoms are as stable as in GR.  

\section{Resolution of Cosmic Singularities}

The unexpected effects of the Palatini gravitational interaction for infrared corrected models in atomic systems 
raises a number of natural questions. What is the physical mechanism responsible for the atomic disintegration? Where does that energy come from? Understanding these points is very important because in the atomic system studied above the gravitational field has been able to surpass the electrostatic forces that keep the electron bound to the nucleus. This means that we are facing a new gravitational mechanism able to generate strong repulsive forces even when tiny amounts of matter and energy are involved. It should be noted that such forces arise in regions where the gradients $\partial_\mu f_R$ and $\nabla_\mu\nabla_\nu f_R$ are strong. In fact, it is precisely those terms which represent the maximum contribution to the Hamiltonian through the induced potential $V_\Omega$ and which introduce strong non-perturbative effects in the atomic dynamics. Thus, the rapid variation of the terms $\partial_\mu f_R$ and $\nabla_\mu\nabla_\nu f_R$ in regions where the characteristic scale of the Lagrangian is reached generates important contributions on the right hand side of the metric field equations (\ref{eq:Gmn}) and results in a strong backreaction. The question now is to see if such repulsive forces and backreaction effects can be achieved in other scenarios of interest. We will show next that ultraviolet corrected models may also give rise to very strong repulsive forces in critical situations such as in the very early Universe. This will confirm that Palatini $f(R)$ theories possess the necessary ingredients to get rid of the Big Bang singularity in quite general situations. \\

\begin{figure}
\includegraphics[width=0.55\textwidth]{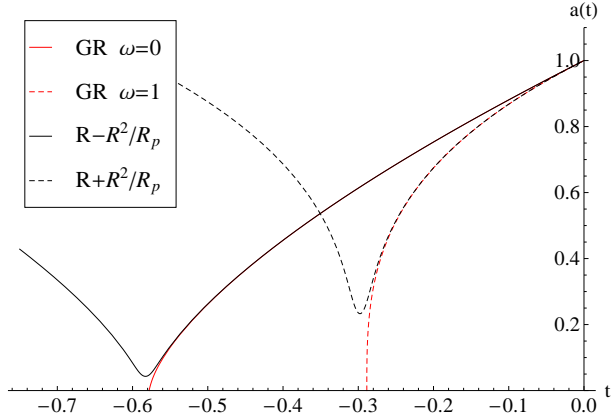}
\caption{Bouncing cosmologies with $K=0$ compared with the GR solution (red lines). The continuous black curve represents the case $w=0$, corresponding to $f(R)=R-R^2/|R_P|$. The dashed line represents the case $w=1$ of the theory $f(R)=R+R^2/|R_P|$. The initial density is the same for the four curves. \label{fig:K0}}
\end{figure}

As an illustrative example, let us consider the following modified Lagrangian $f(R)=R+R^2/R_P$, where $R_P$ represents a curvature scale which could be identified with the inverse Planck length squared. For this Lagrangian we find that $\R=-\kappa^2T$ and $f_R=1-2\kappa^2T/R_P$. A glance at the field equations (\ref{eq:Gmn}) indicates that the right hand side is dominated at low curvatures by the usual Einstein term $\kappa^2T_{\mu\nu}$, with corrections of the order $\sim \kappa^2T/R_P$ which only become important at very high curvatures. At low curvatures/densities, the dynamics is thus very well described by GR. It is important to note that the terms $\partial_\mu f_R  \partial_\nu f_R$ are of order $\sim (\kappa^2\partial_\mu T/R_P)^2$ and, therefore, do not contribute to the first-order corrections. We will return to this point shortly.  \\

When particularized to FLRW cosmologies for matter with constant equation of state $P=w\rho$, the time-time component of the field equations (\ref{eq:Gmn}) leads to 
\begin{equation}
H^2=\frac{\R}{(1-3w)}\frac{\left(1+\frac{2\R}{R_P}\right)\left(1+\frac{1-3w}{2}\frac{\R}{R_P}\right)}{\left[1-(1+3w)\frac{\R}{R_P}\right]^2} \label{eq:H} 
\end{equation}
where $H\equiv \dot a/a$, and we have set $K=0$ for simplicity. From this equation, one readily sees that $H^2$ vanishes when $f_R\to 0$, which corresponds to $\R=-R_P/2$ (see \cite{BOSA09} for a thorough discussion of the conditions at which $H^2$ can vanish). If $R_P>0$, this condition is satisfied by sources with $w>1/3$. If $R_P<0$, then the condition is satisfied by $w<1/3$. The vanishing of $H^2$ in this model occurs when $\ddot a>0$ and implies that the expansion factor has reached a minimum. This means that the Universe started in a contracting phase, reached a minimum size, and then bounced to a new expanding phase, as shown in Fig.\ref{fig:K0}. From this plot one can verify that at low curvatures the expansion factor follows very accurately the GR solution. It is just very near the characteristic curvature/density  scale of the theory where the solution departs from GR and the singularity is avoided. The reason for this dynamical change can be found in the non-perturbative effects introduced by the terms $\nabla_\mu\nabla_\nu f_R$ and $\partial_\mu f_R  \partial_\nu f_R$ on the right hand side of (\ref{eq:Gmn}), which are multiplied by $f_R^{-1}$ and $f_R^{-2}$, respectively. Though, as we advanced above, the latter contributions do not play any role at the perturbative level, they dominate the right hand side of the equations and lead to a strong backreaction that cures the singularity when the characteristic curvature/density scale of the theory is reached. We thus see that in ultraviolet modified models there exist non-perturbative mechanisms able to protect the theory from certain singularities. Whether or not these mechanisms appear in Universes with less symmetry or in other critical situations such as in black hole scenarios is a matter which deserves further study.

\section{Summary and conclusions}  

In this work we have studied two generic Palatini $f(R)$ models: one modified in the infrared, $f(R)=R-\mu^4/R$, and another modified in the ultraviolet, $f(R)=R+R^2/R_P$.  The infrared corrected model, initially considered as a potential candidate to explain the cosmic speedup, has been shown to be ruled out by observations. Models like this one are characterized by infrared curvature corrections which cause strong backreaction effects in regions of very low matter-energy density such as the zeros of atomic wavefunctions. We have shown that the ground state of Hydrogen can be studied using perturbative methods, which guarantees the validity of our results, and turns out to be unstable. Other excited states exhibit non-perturbative behavior and confirm that strong modifications of the physics at such scales should occur, which supports our conclusions. \\ 
We would like to remark that the $1/R$ model in Palatini formalism had not been convincingly ruled out until the publication of \cite{Olmo08a}. This model, first proposed by Vollick \cite{Vol03} in Palatini formalism, was immediately criticized by Flanagan \cite{Fla04a} arguing that it could be ruled out according to our knowledge on electron-electron scattering experiments. Despite Flanagan's efforts  to carry out explicit calculations to verify that claim and respond to Vollick's comment \cite{Vol04}, he concluded \cite{Fla04b} that strong coupling effects could make impossible the calculation of observable physical effects at experimentally accessible energies. The theory could thus not be ruled out on grounds of electron-electron scattering. With the analysis of the Hydrogen atom presented here \cite{Olmo08a}, whose ground state can be treated fully perturbatively, we believe that this discussion has been finally settled. \\

On the other hand, ultraviolet corrected models manifest interesting features in the context of the very early Universe. Non-perturbative mechanisms, analogous to those that spoil infrared corrected models, act in this case but with a successful result: the Big Bang singularity can be removed. In the cuadratic model considered here, the solution of the modified equations represents a deformation of the general relativistic solution which depends on the parameter $R_P$. (This is so because in Palatini $f(R)$ there are no new degrees of freedom and the equations are of second-order, like in GR). If this parameter is set to infinity, then the singular solution of GR is recovered. For any finite value of $R_P$ the cosmic singularity is avoided\footnote{The quadratic Lagrangian $f(R)=R+R^2/R_P$ has been criticized in \cite{BSM} on the basis of the existence of surface singularities in spherically symmetric stellar solutions. However, for Planck scale corrected models such singularities could never exist in nature (see \cite{Olmo08b} for details). } (for some equations of state). Furthermore, the avoidance of the singularity is a non-perturbative effect which could not be guessed from a perturbative analysis at low energies. This result is rooted in the new role played by the matter sources on the geometry due to its coupling to the metric via the connection. We thus see that Palatini $f(R)$ theories provide new gravitational mechanisms that increase the role of the matter sources in the spacetime dynamics. \\ 

Though Palatini $f(R)$ theories were originally motivated by phenomenological considerations (cosmic speedup), today we know that they may be suitable for describing certain properties of quantum spacetimes. In fact, it was recently discovered \cite{Olmo-Singh} that the effective equations of loop quantum cosmology \cite{LQC}, an approach to quantum cosmology based on non-perturbative quantization techniques, can be derived from a Palatini $f(R)$ theory with high curvature corrections. This theory leads to a non-singular bouncing cosmology, and the mechanism responsible for the bounce is analogous to that acting in the $R+R^2/R_P$ model considered here. Our results support the need for further analyses of the dynamics of Palatini theories.

\begin{theacknowledgments}
The author thanks MICINN for a Juan de la Cierva contract, the Spanish Ministry of Education and its program ``José Castillejo'' for funding a stay at the CGC of the University of Wisconsin-Milwaukee, and the Physics department of the UWM for their hospitality during the elaboration of this work. The author's work has also been partially supported by grant FIS2008-06078-C03-02. 
\end{theacknowledgments}


\begin{thebibliography}{40}

\bibitem{book1} Weinberg S., \textit{Cosmology}, Oxford University Press, (2008).

\bibitem{book2}
V. Mukhanov, \textit{Physical Foundations of Cosmology}, Cambridge University Press, (2005).

\bibitem{book3}
Dodelson S., \textit{Modern Cosmology}, Academic Press, (2003).

\bibitem{Capozziello:2009nq}
  S.~Capozziello, M.~De Laurentis and V.~Faraoni,
  arXiv:0909.4672 [gr-qc].
  
\bibitem{Sotiriou:2008rp}
  T.~P.~Sotiriou and V.~Faraoni,
  arXiv:0805.1726 [gr-qc].

\bibitem{Novello:2008ra}
  M.~Novello and S.~E.~P.~Bergliaffa,
  Phys.\ Rept.\  {\bf 463}, 127 (2008)
  [arXiv:0802.1634 [astro-ph]].


\bibitem{Olmo05}
G.J. Olmo, {\it Phys. Rev. Lett.} {\bf 95}, 261102 (2005); {\it Phys.Rev.} {\bf D} 72, 083505 (2005).

\bibitem{Olmo07b}
G.J. Olmo, {\it Phys.Rev.} {\bf D} 75,(2007)023511, gr-qc/0612047. 

\bibitem{Parker80PRL}
L. Parker, {\it Phys. Rev. Lett.} {\bf 44}, 1559 (1980).

\bibitem{Parker80}
L.Parker, {\it Phys.Rev.} {\bf 22}, 1922 (1980)

\bibitem{P-P82}
L.Parker and L.O. Pimentel, {\it Phys.Rev.} {\bf 25}, 3180 (1982)

\bibitem{Olmo08a}
  G.~J.~Olmo,
  Phys.\ Rev.\  D {\bf 77}, 084021 (2008)
  [arXiv:0802.4038 [gr-qc]].

\bibitem{Olmo07}
G.J. Olmo, {\it Phys. Rev. Lett.} {\bf 98}, 061101 (2007).  
 
\bibitem{CDTT04}
S.M. Carroll, V. Duvvuri, M. Trodden and M.S. Turner, 
{\it Phys. Rev.} {\bf D} 70, 043528 (2004).

\bibitem{Vol03}
D.N. Vollick, {\it Phys. Rev.}{\bf D} 68, 063510 (2003), astro-ph/0306630.

\bibitem{BOSA09}
  C.~Barragan, G.~J.~Olmo and H.~Sanchis-Alepuz,
  Phys.\ Rev.\  D {\bf 80}, 024016 (2009)
  [arXiv:0907.0318 [gr-qc]].

\bibitem{Fla04a}
E.E.Flanagan, {\it Phys.Rev.Lett.}{\bf 92}, 071101 (2004).

\bibitem{Vol04}
D.N. Vollick, {\it Class.Quant.Grav.} {\bf 21}, 3813 (2004).

\bibitem{Fla04b}
E.E.Flanagan, {\it Class.Quant.Grav.} {\bf 21}, 3817 (2004).

\bibitem{BSM}
E. Barausse, T.P. Sotiriou, and J.C. Miller, {\it Class.Quant.Grav.} 25, 062001  (2008); {\it Class.Quant.Grav.} 25, 105008 (2008).


\bibitem{Olmo08b}
G.J.Olmo, {\it Phys.Rev.} {\bf D} 78,104026 (2008).

\bibitem{Olmo-Singh}
G.J. Olmo and P. Singh, JCAP 0901,  030 (2009).

\bibitem{LQC}
M. Bojowald, {\it Living Rev. Rel.} {\bf 8}, 11 (2005);
A. Ashtekar, T. Pawlowski and P. Singh, {\it Phys. Rev. Lett. } {\bf 96}, 141301 (2006); {\it Phys. Rev.} {\bf D} 73,  124038 (2006);
{\it Phys. Rev. } {\bf D} 74, 084003 (2006).

\end{thebibliography}
\end{document}